\documentclass{vgtc}                          

\ifpdf
  \pdfoutput=1\relax                   
  \usepackage{graphicx}                
  \DeclareGraphicsExtensions{.pdf,.png,.jpg,.jpeg} 
\else
  \usepackage{graphicx}                
  \DeclareGraphicsExtensions{.eps}     
\fi%

\graphicspath{{figures/}{pictures/}{images/}{./}} 

\usepackage{microtype}                 
\PassOptionsToPackage{warn}{textcomp}  
\usepackage{textcomp}                  
\usepackage{mathptmx}                  
\usepackage{times}                     
\usepackage{cite}                      
\usepackage{tabu}                      
\usepackage{booktabs}                  
\usepackage{tabularx}
\usepackage{ragged2e}
\usepackage{multirow}
\usepackage{enumitem}

\usepackage[dvipsnames]{xcolor}
\usepackage[normalem]{ulem}
\definecolor{mred}{rgb}{.80,.12,.30}
\definecolor{MRED}{rgb}{.80,.12,.30}
\definecolor{grey}{rgb}{0.5,0.5,0.5}
\definecolor{purple}{rgb}{.75,0,.85}
\definecolor{pistachio}{rgb}{0.58, 0.77, 0.45}

\newif\ifnotes
\notestrue 

\onlineid{1321}

\vgtccategory{Research}

\vgtcinsertpkg

\title{Facilitating Reproducibility in Visualization Systems Research Through Containerization}

\author{Thomas Breimer\thanks{e-mail: tbreimer@brandeis.edu}\\ %
        \scriptsize Brandeis University %
\and Pirm Dhararag\\ %
        \scriptsize Brandeis University %
\and Kiya Gedamu\\ %
        \scriptsize Brandeis University %
\and Paras Pandey\\ %
        \scriptsize Brandeis University %
\and Marie Amandine Reyes\\ %
        \scriptsize Brandeis University %
\and Tal Spector\\ %
        \scriptsize Brandeis University %
\and Galileo Steinberg\\ %
        \scriptsize Brandeis University %
\and Nathan Subrahmanian\\ %
        \scriptsize Brandeis University %
\and Paul Weir\\ %
        \scriptsize Brandeis University %
\and Peter Williams\\ %
        \scriptsize Brandeis University %
\and Weichen Zhao\\ %
        \scriptsize Brandeis University %
\and Dylan Cashman\thanks{e-mail: dylancashman@brandeis.edu}\\ %
     \scriptsize Brandeis University %
     }

\abstract{In this position paper, we argue for the importance of containerizing visual analytics artifacts to improve the scientific progress of our research community.  First, based on our previous experience authoring, reading, and reviewing visual analytics research from the previous decade, we identify risks and impediments to empirical research in visual analytics within the broader reproducibility crisis, including an inability to compare systems and techniques and a slow pace of adjusting to a rapidly changing ecosystem of analysis methods, such as machine learning and artificial intelligence.  Next, we analyze the benefits of different artifact sharing and deployment setups across five artifact goals: reproducibility, replicability, portability, performance, and scalability.  Then, we argue that containerization addresses the issue of reproducibility in visual analytics research, with downstream benefits in terms of replicability, portability, performance, and scalability. To demonstrate the ease of use and the utility of containerization, we deploy ten visual analytics systems developed over the course of a graduate seminar onto a live server using containerization with the help of agentic AI.  We share lessons learned through the containerization process, and describe a suite of potential benefits for our research community as well as research opportunities for leveraging modularity to improve the artifacts themselves.} 


\begin{document}



\teaser{
  \centering
  \includegraphics[width=\linewidth]{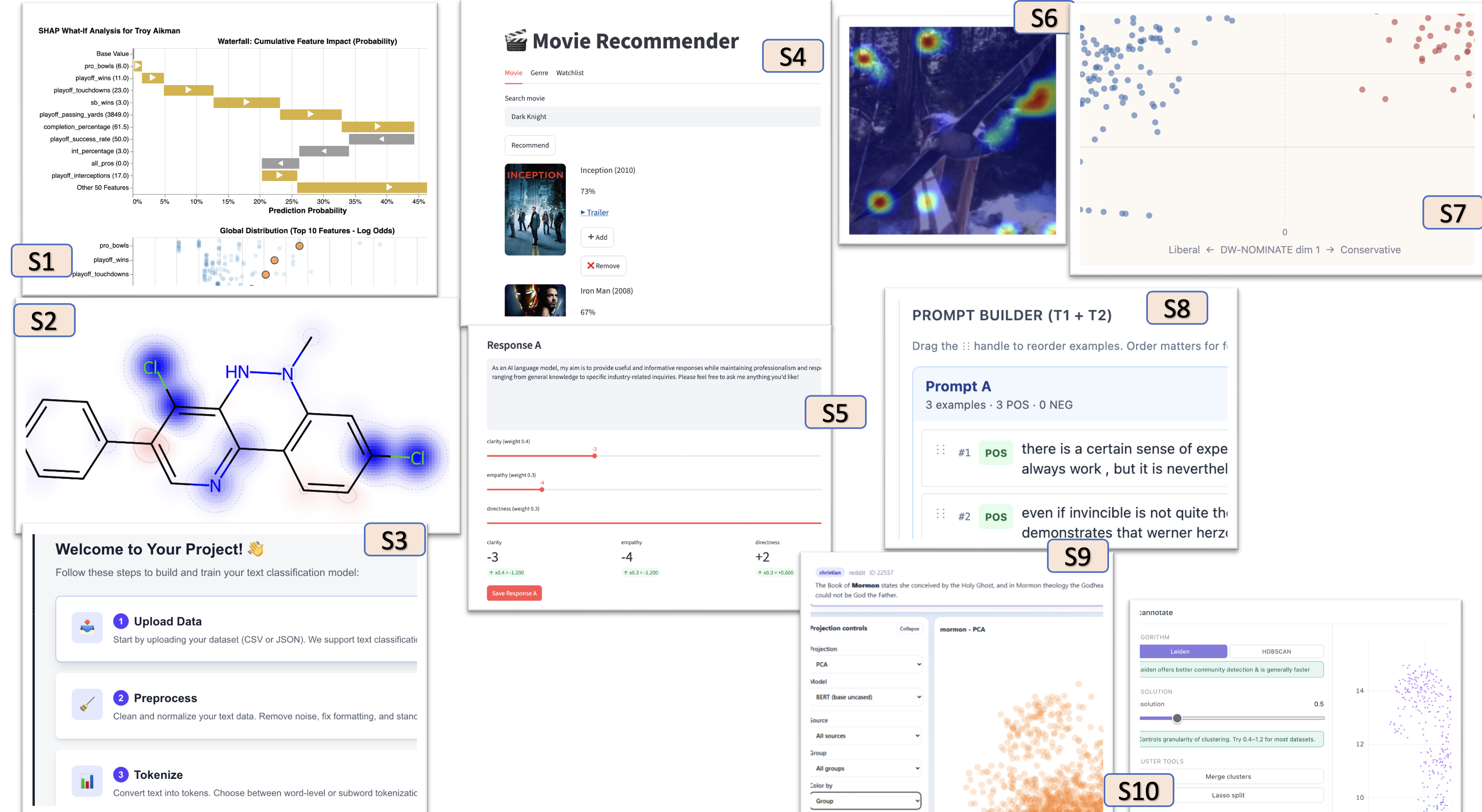}
  \caption{Ten interactive analytics systems designed as part of an interactive machine learning graduate seminar.  All ten systems were made available through docker containers and are served at a public website.  Through these applications, we demonstrate the use of containerization to increase reproducibility in visualization systems research.  System labels reference Table~\ref{tab:systems}.}
  \label{fig:systems}
}

\maketitle

\section{Introduction}


Visual analytics (VA) research often centers around the development of artifacts, such as systems, individual components, and algorithms that drive those components and systems.  The focus on artifacts, however, leads to some complicating factors in building epistemic knowledge.  In our experience, the three most significant factors are 1) the \textit{lack of controlled variables of interest} that represent the phenomenon being studied, 2) the \textit{high computational and temporal cost of development and deployment of artifacts} that exist within ecosystems of technology that is constantly changing, and 3) the \textit{brittle nature of deployment of these artifacts}, complicating comparison with existing methods to enable artifacts to build upon assumptions and produce empirical authority on the best methods.

We believe that the first factor is the focus of much existing research developing hybrids of quantitative and qualitative methods~\cite{wall2018heuristic, suh2022inferential, ceneda2023heuristic, islam2024strategies}.  However, the second two factors are relatively understudied and reflect a threat to reproducibility.  If it were easier to rebuild previous artifacts and reproduce findings, it would be more common for artifacts to be incorporated into comparative studies and larger metaanalyses of broad open problems in visual analysis.  This could potentially enable evaluation platforms such as ReVISit 2~\cite{cutler2025revisit2}.  Similarly, if it were easier to replicate functionality from artifacts, it could speed up development of future artifacts to enable greater engagement with other fast-moving fields such as machine learning and artificial intelligence research.  Thus, we argue that addressing the understudied problems of the high cost of development and the brittle nature of resulting artifacts has potential for high impact in improving the quality of our science and its impact.

We proffer a simple technical palliative to these two factors: \textbf{containerization}~\cite{moreau2023containers}. Containerization is an application deployment paradigm where all application dependencies, libraries, and a suitable execution environment is packaged alongside the app's source code. Unlike virtual machines, which require a substantial overhead due to the execution of a ``guest OS" concurrently with the native operative system, a container shares the host machine's kernel and is encapsulated via a dedicated user space. Thus, one container may run on a variety of different machines and architectures with little additional development work or computational overhead. For example, consider a visual analytics application which uses XAI techniques to predict molecule properties \textit{(MolExplain (\textbf{S2}), discussed later in this work)}. Such a system requires computationally intensive Python-based XAI algorithms to produce insights, which may be presented via a Javascript-based web frontend compiled using a modern development server like Vite\footnote{https://vite.dev}. Thus, even if all the code is readily available, anyone interested in running MolExplain locally would be faced with a tedious setup process including initializing both Python \& NodeJS development environments. However, if MolExplain was containerized, these environments would be packaged with the container, avoiding a drawn-out setup process and facilitating reproducibility.

In this process of containerization, the ecosystem in which a research artifact is developed is used to define a portable \textit{container} that can run that same artifact on most other hardware (i.e. personal laptops, servers, cloud infrastructure, etc.).  Compared to the code alone, providing a container makes it trivial to reproduce the artifact in a running state.  In addition, with some conventions that could be codified in a schematic format, containerized visual analytics techniques could be composed in a plug-and-play, modular format.  For example, MolExplain (\textbf{S2}) could be abstracted into two separate containers which communicate via a local network; the Python backend which generates explainability data, and the Javascript frontend which displays it to the user.  If a new technique for molecule visualization or XAI presentation were proposed, and existing techniques all were available as containers in a standardized format, this new technique could be rapidly evaluated against existing techniques, and if found superior in some contexts, could be switched out in deployed tools.  Containerization potentially has additional downstream benefits for research code, impacting portability and scalability as well.  In this paper, we posit that containerized artifacts compare favorably with six other possible architectures for research artifacts along these dimensions of utility.

Containerization itself is not novel and has been a popular technique in DevOps and fields burdened by complex, dependency-heavy computational workflows, such as bioinformatics and high-performance computing (HPC) ~\cite{osti_1627824, 10.3389/fbinf.2021.757291}. However containerization is rarely used for sharing code from visual analytics artifacts.  It can be difficult to manually define the containers, as container technologies have their own domain specific languages that rely on a sophisticated understanding of operating systems and package management.  However, recent software development tools using AI agents have rendered the process trivial.  To demonstrate this, we describe our experience taking 10 independently developed visual analytics systems, developed as part of a graduate seminar on Interactive Machine Learning Systems, and containerize them using an agentic AI assistant.  We are able to complete each containerization in an average of 30-60 minutes.  Once containerized, each application is able to be deployed to an online server with minimal effort, and the resulting systems can be viewed at \url{beliv.viscontainers.org}. This website will also feature a form for researchers to submit their VA systems for containerization as part of a future research project. We argue that releasing a runnable container could be lightly encouraged through recognition akin to reproducibility stamps~\cite{isenberg2024reproducibility}. We also speculate about potential research directions once it becomes normalized to release artifacts with runnable containers.  For example, identifying the design space of communication between different types of artifacts could illuminate the role of particular components of systems; for instance the role a configuration component plays versus the tabular view of the data or visual components like scatterplots.






In summation, this position paper contributes the following:
\begin{itemize}
    \item A comparison weighing the pros and cons of different visualization system architecture methods for many utilities including reproducibility, replicability, portability, performance, and scalability.
    \item An easy-to-follow outline of containerizing visual analytics artifacts, demonstrated through a case study of containerizing and hosting ten novel visual analytics systems.
    \item A request for readers to send the authors their visualization systems to facilitate further research.
\end{itemize}

\section{Visual Analytics Artifacts}

Of the seven types of research contributions in Human Computer Interaction identified by Wobbrock and Kientz in 2016~\cite{wobbrock2016research}, visual analytics research frequently results in artifact contributions, which are typically bundled and distributed as part of systems.  Consider the task of \textit{model steering}, in which a human-in-the-loop exercises some control over the training of a machine learning model.  Visual analytics systems for model steering have been developed for many different types of models (i.e. regression models~\cite{das2019beames}, classification models~\cite{kapoor2010interactive, tamagnini2017interpreting}, deep learning models~\cite{cashman2019ablate}, dimensionality reduction models~\cite{jeong2009ipca}, clustering models~\cite{cavallo2018clustrophile})



Wobbrock and Kientz identify three types of artifacts as HCI contributions: systems, techniques, or designs.  But in visual analytics research, even if the contribution is a technique or a design, it is typically incorporated into a system in order to demonstrate (or instantiate) it.  This might be a necessity of an evaluation, or to show the feasibility of the technique so that it can be used to drive use cases.  A side effect of this expectation is that the actual artifact is either a system or a technique/design that is tightly coupled with the system that is made out of necessity.  This is not always the case; the \textit{draco} recommendation engine and its sequel \textit{draco2} are offered as standalone modules that can be integrated into systems separate from the infrastructure that they used in their use cases and evaluations~\cite{moritz2018formalizing, yang2023draco}.  However, artifacts are not usually so easily isolated and shared, and the continued maintenance of software takes significant development time. Draco2, for example, has received over 1000 commits to its Github repository since 2020 at the time of writing. Typically, it's difficult to run and maintain research artifacts from even a few years earlier, and the continued maintenance of these artifacts takes significant time which eats into academic timelines.

\subsection{Existing Frameworks of Artifact-based Research}

Existing work has sought to define what a research artifact is in the context of visualization and enable the integration of artifacts into the wider visualization research ecosystem. Andrienko et al. presented an Artifact-Transform Workflow Language (ATWL)~\cite{andrienko2026atwl}, which leverages a modular ontology of eight artifact types to describe VA workflows and demonstrated the ability of LLMs to extract workflows from published visualization work. Work by Ferreira et al. constructed a modular framework for describing and building urban VA systems~\cite{ferrerira2026vablueprint}, providing a modular approach to constructing domain-specific VA systems, also leveraging LLMs to ground techniques in published visualization work. To achieve a more domain-agnostic VA system architecture, Heer and Moritz presented Mosaic, an architecture for interactive visualization systems which decouples data processing and presentation~\cite{mosaic-arch}. While these works facilitate the composability of VA workflows and systems, the distribution and continued operability of artifact-containing systems is crucial for extending existing research. Isenberg evaluated to what extent different architectures facilitate reproducibility for visualization research prototypes~\cite{isenberg2022experiences}, noting Windows Executables as a particularly salient architecture and advocating for increased adoption of open science practices to facilitate reproducibility.

\subsection{Limitations of Artifact-based Research}

The bias towards systems as artifact results in some risks to both the scientific and broader impact of visual analytics research. 

\noindent \textbf{C1: It is difficult to reproduce findings for metaanalyses or comparisons.}  If artifacts are tightly coupled and dependent on many different packages and the researcher's local development environment, it can be hard to run them once that environment changes.  This means that reproducibility capabilities degrade rapidly as time goes on.  As a result, it becomes difficult to collect multiple artifacts to conduct further analysis or compare techniques.  For example, in order to compare interactive clustering systems, you would need to get many disparate systems running on disparate hardware. 

\noindent \textbf{C2: Artifacts cannot be applied to new datasets to increase impact.} Again, if artifacts are hard to get running, then they will not be adopted by the target users.  Many research contributions consider visualization designers as one of the target audiences.  However, outside of viewing static images within the paper, it can be hard to understand how an artifact will function outside of the use cases presented in its corresponding publication. 

\noindent \textbf{C3: Systems are composed of too many uncontrolled variables.}  If a contribution is a particular technique, but it is evaluated in the context of a system which is built anew, it results in many uncontrolled variables.  For example, if a new technique for interactive cluster construction is proposed, and evaluated with a system built from the ground up that includes many features (scatterplot with semantic zoom, semantic interactions, recommendations, etc.), it's hard to disentangle the effect of these features from the technique being proposed.

\noindent \textbf{C4: Developing visual analytics systems is expensive in time and expertise.}  Since previously built systems typically are difficult to get running, researchers often have to develop their own features with each new project.  This means that new systems take a long time to develop and are sometimes only accessible to large well-funded research groups.

\subsection{Replication Crisis}


We particularly highlight \textbf{C1}, challenges in reproducibility, as a motivation to consider a new method for sharing artifacts. This rapid decay of software accessibility does not happen in a vacuum; rather, it directly intersects with the broader replication crisis that has become an escalating area of concern across the entire computer science landscape\cite{cockburn2020replication}. A fundamental hurdle in addressing this systemic vulnerability is that replication and reproducibility can be difficult to formally define, with various subfields operating under conflicting definitions of what constitutes a successful replication of data, code, or experimental setup\cite{plesser2018randr}. Within the visualization domain, these broader scientific conversations have led to foundational questions regarding the long-term validity and empirical reliability of visualization research\cite{kosara2016sand}. When visualization systems are treated as prototypes that cannot be re-executed a few years after publication, verifying their scientific claims becomes impossible. 

In response, the visualization community has mobilized, establishing recommendations and guidelines aimed at mitigating the replication crisis within our field\cite{kosara2018replication, franke2023reproducible, haroz2018open, fekete2020reproducibility}. These frameworks advocate for a cultural pivot toward open science, structural transparency, and public data sharing. To institutionalize these values, researchers have adapted concrete strategies from other scientific disciplines. Prominent among these are experimental preregistration\cite{besancon2024merits}, which limits p-hacking and publication bias in empirical user studies, and the deployment of formalized transparency or reproducibility stamps during the peer-review process\cite{isenberg2024reproducibility}. While these interventions have drastically improved the rigor of statistical analysis and evaluation methodologies in visualization, a critical gap remains regarding \textit{systems-level} artifacts. Preregistration can protect an experimental design, and a generic badge can confirm code availability, but neither guarantees that a highly heterogeneous visual analytics system will compile or run across changing hardware ecosystems. This paper builds directly upon these existing community initiatives by shifting the focus toward architectural encapsulation, addressing the technical dimensions of the visualization replication crisis through automated containerization.

\section{Architecture Options for Visualization Systems}


To facilitate the dissemination of new techniques and knowledge in the visualization community, it is necessary for systems to be shared. However, researchers may choose particular computational architectures based on experimental or performance constraints which have additional overhead from standard software engineering requirements. Consequently, visualization systems are bundled with code that is not inherent to the contribution, but may be necessary to serve a user interface, store data, interface between components, or access hardware components. The constraints imposed by these implementation decisions dictate additional barriers to the future viability and usefulness of the system, which impede usefulness in the visualization research ecosystem. In this section, we enumerate seven architecture options used in visualization (Table~\ref{tab:architecture_matrix}), and discuss their advantages and limitations. We argue that containerization facilitates reproducibility with downstream benefits in terms of replicability, portability, performance, and scalability.




\subsection{Web Based Architectures}


The traditional Thin Client-Server Architecture \textbf{(A1)} is the most common paradigm for visual analytics systems today. With this architecture, a lightweight user interface communicates with a backend. The browser-based frontend is typically written in JavaScript or TypeScript using rendering libraries like D3, React, or WebGL. The backend is often powered by Python (Flask\footnote{\url{https://flask.palletsprojects.com/}}, FastAPI\footnote{\url{https://fastapi.tiangolo.com/}}, Django\footnote{\url{https://www.djangoproject.com/}}) or C++, handling computationally demanding tasks and communicating with the client via REST APIs.

To alleviate the need for a heavy-duty server, the Thick Client Server Architecture \textbf{(A2)} compiles high-performance backend logic directly into binary WebAssembly (WASM) instructions, which can be executed by a browser at near-bare-metal speeds. This technique is used by Mosaic \cite{mosaic-arch}, which leverages DuckDB\footnote{\url{https://duckdb.org/}}-WASM to handle client-side queries, and Tensorflow.js \cite{smilkov2019tensorflow}, which uses WASM and WebGPU for client-side machine learning model training. 

To avoid server-side development completely, the Thin Client + SaaS (Software as a Service) Architecture \textbf{(A3)} offloads computation entirely onto a third-party server, sending back information which is displayed on the client side. While this paradigm drastically limits the flexibility of the system, it may be preferred for use cases where client-side computation is infeasible or data is proprietary; for instance workflows which involve Large Language Models.

\subsection{Desktop Executables}

Software compiled directly for specific desktop operating systems offers high-performance execution by bypassing browser abstraction layers, granting direct access to native system memory and hardware-accelerated GPUs. Desktop Executables \textbf{(A4)} are typically implemented using application frameworks like C++/Qt or Java/Swing. This architecture is vital for high-throughput scientific visualization tools like ParaView \cite{ayachit2015paraview} and immersive analytics platforms leveraging virtual and augmented reality, such as the Unity-based toolkit IATK \cite{cordeil2019iatk}. From a preservation standpoint, native executables target single-platform runtimes (e.g., Microsoft Windows), which can offer multi-decade backward compatibility or run reliably inside isolated virtual machines. For this reason, MS Windows™ executables have been suggested as a desirable architecture for visualization research prototypes \cite{isenberg2022experiences}. However, compiling across heterogeneous target platforms (macOS, Linux, x86 vs. ARM) remains an expensive, manually intensive engineering task.


\subsection{Notebooks and Business Intelligence Software}

While less common for bespoke, multi-view visual analytics systems, Computational Notebooks \textbf{(A5)} and Business Intelligence (BI) Software \textbf{(A6)} offer rapid prototyping alternatives that excel in ease of development. Notebook environments, such as Jupyter and Observable, interleave text cells with reactive blocks of executable code. This linear execution format promotes inline visual data analysis and collaborative dissemination via cloud platforms like Google Colab. The Javascript-based Observable Environment \cite{kosara2023notebooks} supports heavily interactive data-driven analysis tasks, while Python-based Jupyter Notebooks are suited for data processing and machine learning workflows, and are a preferred mode of submission for the IEEE Visualization Explainable AI Workshop (visxai)\footnote{\url{https://visxai.io/}}. Conversely, BI Software \textbf{(A6)} provides low-code, declarative interfaces for data exploration. Though highly restricted in interface layout and novel visual encodings, platforms like Tableau are fundamentally grounded in foundational visual analytics research \cite{mackinlay2007tableau}.

\subsection{Containerization with Docker}

Containerization offers a fundamentally different approach to application distribution. Instead of an application resting on an operating system's application layer or within the browser sandbox, a containerized application comes packaged with the environment it needs to run. This package includes minimal operating system utilities, filesystem structures, and dependencies, along with the application code itself. Unlike virtual machines, the containerization paradigm abstracts only the software layer, sharing the host OS kernel. This dramatically reduces the computational resource overhead, making it feasible for many containerized applications to run on one machine at once. A number of different containerization platforms exist, such as Singularity\footnote{\url{https://sylabs.io/singularity/}}, Podman\footnote{\url{https://podman.io/}}, and Docker\footnote{\url{https://www.docker.com/}}. We focus on Docker due to its maturity as a platform, support for multiple operating systems (Linux, Mac, Windows), and potential for scalability \cite{moreau2023containers}. 

Application source code containerized via Docker will come packaged with a \texttt{Dockerfile}, a plain text file containing step-by-step instructions for how to assemble the software environment required for the application. When a \texttt{Dockerfile} is ``built", Docker compiles the application into an \textit{Image}: a read-only, executable template generated by the Dockerfile that includes the application code, libraries, and dependencies required for the application to run. A \textit{Container} is a live, running instance of a Docker image whose execution is facilitated by the \textit{Docker Daemon}. The \textit{Docker Daemon} is a background process which executes containers, isolates them from one another, and enables communication between containers and the host system via port mapping. Docker has been a staple of DevOps for at least a decade, yet little visualization research has leveraged the technology: out of 3,291 journal articles published in IEEE Transactions on Visualization and Computer Graphics from 2021 to 2026, only 21 papers mention ``Docker" (0.64\%), and only 14 papers explicitly specify ``Docker container" (0.43\%)\footnote{Query 1 (Baseline): (``Publication Title": ``Transactions on Visualization and Computer Graphics"). Query 2 (Broad Match): (``Publication Title": ``Transactions on Visualization and Computer Graphics") AND (``Document Text": ``docker"). Query 3 (Phrase Match): (``Publication Title": ``Transactions on Visualization and Computer Graphics") AND (``Document Text": ``docker container").}.


\section{Implications of Architecture Selection}

\begin{figure*}[t]
    \centering
    \includegraphics[width=0.85\linewidth]{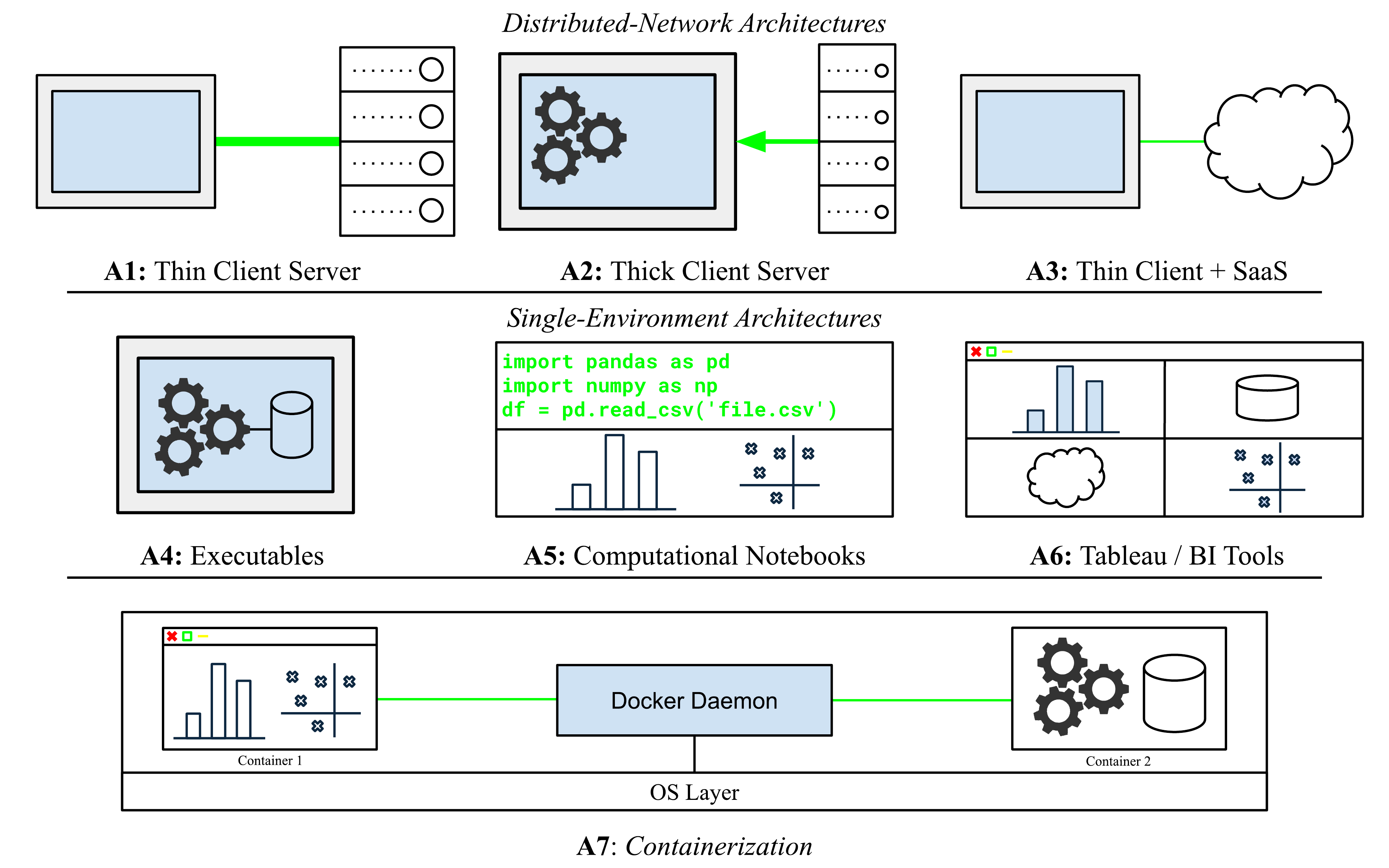}
    \caption{Structural topologies, environment boundaries, and data flow pipelines across the seven visual analytics architecture paradigms (\textbf{A1}--\textbf{A7}), complementing the matrix in Table~\ref{tab:architecture_matrix}. Component wireframes denote the physical separation between user interfaces, execution engines, and data storage layers. Traditional distributed and single-environment paradigms (\textbf{A1}--\textbf{A6}) exhibit high environmental coupling, leaving components vulnerable to local package drift, configuration friction, and host-level software decay over time. In contrast, the containerized paradigm (\textbf{A7}) introduces an isolated runtime encapsulation layer that wraps the application code, abstracting the multi-tiered software stack away from mutable host OS and kernel drivers. The primary takeaway illustrates the fundamental engineering trade-off: containerization maximizes long-term reproducibility and cross-platform portability by intentionally shifting the maintenance burden from individual application software layers to the underlying systems infrastructure.}
    \label{fig:systems_architecutre}

\vspace{-5mm}
\end{figure*}

For each architecture option, we consider the following goals: reproducibility, replicability, portability, performance, and scalability. While reproducibility remains the core goal of this position paper, we posit that each of these qualities are desirable to ensure the future utility and applicability of a system. We score each option according to these metrics as (\textit{low}, \textit{medium}, or \textit{high}), to the extent which a goal is enabled by that architecture option. Our determinations are shown in Table \ref{tab:architecture_matrix} and justifications for these ratings are detailed in the subsequent sections. Additionally, Figure \ref{fig:systems_architecutre} provides illustrations of the different architecture options discussed.  

\begin{table*}[t]
\centering
\caption{A comparison of the seven visualization system architecture options \textbf{(A1)}-\textbf{(A7)} evaluating their respective advantages and limitations across five artifact goals: reproducibility, replicability, portability, performance, and scalability. The matrix highlights that while traditional web-based and desktop native solutions optimize for immediate performance and accessibility, they often struggle with multi-year reproducibility without the adoption of isolated environments such as containerization.}
\label{tab:architecture_matrix}
\begin{tabular}{lcccccc}
\toprule
\textbf{Architecture Type} & \textbf{Reproducibility} & \textbf{Replicability} & \textbf{Portability} & \textbf{Performance} & \textbf{Scalability} \\
\midrule
\textbf{A1}: Thin Client Server & Medium & Medium & Medium & Medium & High \\
\textbf{A2}: Thick Client Server  & High & Low & High & High & Low \\
\textbf{A3}: Thin Client + SaaS & Medium & High & High & Low & High \\
\textbf{A4}: Executable & Medium & Low & Low & High & Low \\
\textbf{A5}: Computational Notebooks & Low & High & High & Low & Medium \\
\textbf{A6}: Tableau / BI Tools  & Medium & High & Medium & Medium & High \\
\textbf{A7: Containerized} & \textbf{High} & \textbf{Medium} & \textbf{High} & \textbf{High} & \textbf{High} \\
\bottomrule
\end{tabular}
\begin{flushleft}
\end{flushleft}
\vspace{-10mm}
\end{table*}

\subsection{Reproducibility \& Replicability}

We adapt the ACM's formal definitions of computational reproducibility and replicability. Specifically, an experiment is considered \textit{reproducible} when ``an independent group can obtain the same result using the author’s own artifacts" and \textit{replicable} when ``an independent group can obtain the same result using artifacts which they develop completely independently."\footnote{\href{https://www.acm.org/publications/policies/artifact-review-and-badging-current}{https://www.acm.org/publications/policies/artifact-review-and-badging-current}} Within the context of visualization research, we define an architecture and deployment strategy as reproducible if the resulting artifact can be distributed, installed, and executed with minimal configuration or debugging overhead (e.g., resolving conflicting library dependencies or isolating target runtimes). Conversely, we define a strategy's replicability by the degree to which external researchers can inspect, modify, or extend (``hack") the system's underlying logic without confronting prohibitive technical barriers or specialized domain friction.

Browser-based user interfaces inherently offer a highly accessible solution for distribution. By compiling client-side processing logic directly into WebAssembly modules (\textit{e.g.}, DuckDB-WASM, Pyodide), a Thick Client-Server (\textbf{A2}) architecture can execute entirely within a local browser environment without requiring an active backend server. This client-side self-containment insulates the system from infrastructure failures, which frequently plague traditional Thin Client configurations (\textbf{A1}, \textbf{A3}) post-publication. A parallel failure mode manifests in local Computational Notebooks (\textbf{A5}) and Executables (\textbf{A4}), where ``dependency hell" \cite{moreau2023containers}, occurring via conflicting Python package versions or missing platform-specific dynamic-link libraries (\texttt{.dll}), frustrates long-term reproducibility efforts. While commercial platforms like Tableau (\textbf{A6}) simplify deployment workflows for non-technical audiences, they present barriers to academic reproducibility due to their proprietary licensing structures and version-locked ecosystems.

Conversely, replicability is tightly bound to the learning curve of an architecture's underlying tech stack. Because Computational Notebooks (\textbf{A5}) and BI Tools (\textbf{A6}) are intentionally engineered for rapid prototyping via modular execution cells and declarative interfaces, they achieve high replicability marks. The traditional Thin Client-Server model (\textbf{A1}) introduces moderate replication overhead due to the complexities of developing, orchestrating, and maintaining discrete frontend and backend modules. Thick Client-Server systems (\textbf{A2}) present an even higher barrier, demanding specialized web engineering expertise to compile visual analytics workflows into optimized low-level WebAssembly bytecode. Conversely, the Thin Client + SaaS configuration (\textbf{A3}) lowers frontend barriers by abstracting backend infrastructure entirely, while sacrificing system transparency. Desktop Executables (\textbf{A4}) represent the most significant bottleneck to replicability; compiling source code into native binary payloads renders the underlying implementation virtually opaque and unrecoverable to external researchers.

\subsection{Portability}

Ideally, a visual analytics system should execute across diverse hardware configurations with minimal cross-platform development friction. We define \textit{portability} as the capacity of a system to run across heterogeneous hardware architectures, operating systems, and deployment environments without necessitating substantive modifications to its codebase. Because web browsers serve as standard application runtimes across almost all modern consumer operating systems, Thick Client-Server (\textbf{A2}) and Thin Client + SaaS (\textbf{A3}) architectures exhibit inherently high portability. Standard Thin Client-Server models (\textbf{A1}) achieve slightly lower marks, as shifting between cloud service providers or on-premise institutional servers demands configuration adjustments. Computational Notebooks (\textbf{A5}) offer flexible portability due to their cross-platform compatibility with containerized notebook hubs or managed cloud environments like Google Colab. In contrast, BI Tools (\textbf{A6}) often suffer from fractured platform support: for example, Tableau Desktop lacks native support for Linux. Desktop Executables (\textbf{A4}) present the lowest portability, as binaries are compiled for highly specific CPU architectures and operating system kernels, frequently relying on platform specific libraries. True portability enables the highest level of reproducibility, as anyone wishing to use an artifact for themselves is not restricted to one hardware architecture. However, portability may come at a cost in terms of artifact capability and development time required. 

\subsection{Performance \& Scalability}

Performance and scalability are inextricably linked dimensions that evaluate a system's capacity to process massive data volumes and adapt efficiently to high-power hardware. System \textit{performance} encompasses not only initial data load times, but also continuous interactive attributes such as frame rates, interface responsiveness during brushing-and-linking, and memory consumption. Conversely, \textit{scalability} denotes an architecture's capability to leverage high-powered distributed computing solutions, such as cluster nodes. Since offloading computation to servers distances the user from the execution engine, optimizing for both performance and scalability simultaneously presents a fundamental architectural challenge.

Architectures that execute analytical pipelines directly on local client hardware, such as Thick Client-Server systems (\textbf{A2}) and Desktop Executables (\textbf{A4}), eliminate network latencies. Native executables achieve the highest performance ceiling by running compiled machine instructions directly on host system hardware; however, optimized client-side browser engines running WebAssembly can achieve near-native execution thresholds for Thick Clients (\textbf{A2}). While traditional Thin Client models (\textbf{A1}, \textbf{A3}) ensure fluid visual interactions by keeping the browser layer lightweight, they remain strictly bound by the network latencies required to transfer data across the server-client wire. Managed SaaS backends (\textbf{A3}) are particularly vulnerable to external resource contention and API rate-limiting during peak traffic periods, degrading interactive performance. While Computational Notebooks (\textbf{A5}) can leverage local or remote acceleration for heavy machine learning evaluations, passing data structures from a computational kernel to the browser relies on expensive data serialization processes. This overhead frequently throttles interface throughput and introduces visible interaction lag, though emerging frameworks seek to mitigate this barrier by unifying notebook state with interactive display mechanisms \cite{wu2020b2}. Finally, BI Tools (\textbf{A6}) leverage highly optimized, specialized database extraction and caching engines \cite{wesley2011tableau} to deliver high performance for standard visualization layouts, but their rendering pipelines degrade rapidly when forced to handle non-standard visual encodings or real-time human-in-the-loop steering cycles.

Regarding scalability, server-based backends \textbf{(A1}, \textbf{A3}) present the highest potential, as they decouple compute from client-side physical resource constraints. BI Tools (\textbf{A6}) operating within managed cloud environments leverage elastic SaaS resources to scale operations across distributed servers. Similarly, Computational Notebooks can be run on cloud kernels. Conversely, Thick Client (\textbf{A2}) and Desktop Executable (\textbf{A4}) frameworks offer little horizontal scalability; local executables are restricted by the fixed physical specifications of the user's workstation, while web applications relying on local browser-side execution (\textbf{A2}) are capped by the host browser's security sandbox and memory allocation limits.

\subsection{Why Containerization?}

The discussion in the previous section is characterized by an interlocking set of tradeoffs which limits visualization systems as soon as an architecture is chosen. A developer may want to prioritize UI fluidity and limit network latency, and thus would be drawn to an Executable (\textbf{A4}) or Thick Client (\textbf{A2}) architecture. If portability is a requirement, then this narrows the choice to Thick Client (\textbf{A2}). Thus, the developer is forced to build their system using complicated web technology, limiting future replicability and stretching development time. Containerization provides a solution to these tradeoffs, as packaging an application alongside its dependencies and runtime environment enables maximum reproducibility without compromising on replicability, portability, performance, or scalability. This is because containerization is designed with a portability-first approach, yet does not compromise on performance: research has shown that Docker containers can run at speeds near native application for genomic workflows \cite{ditommaso2015impact}. This eliminates the traditional performance-portability tradeoff, as Docker containers can be run on a variety of different architectures and hardware with little setup.

The reproducibility advantage enabled by containers is facilitated and compounded by their downstream advantages in terms of our remaining artifact goals. Since containers are isolated from one another and communicate through standardized APIs, visualization systems may be modularized. Different containers could store the core contributions of visualization research separately from boilerplate frontend or backend code, allowing future teams to isolate and build off of visualization artifacts. This modularity also emphasizes scalability, as backend components may be run locally during development, then glean the benefits of parallel processing in cloud deployments. Kubernetes\footnote{\url{https://kubernetes.io/}}, in particular, allows for horizontal scaling of containerized applications by facilitating the deployment and management of dozens or hundreds of containers on cloud or physical servers including auto-scaling, storage orchestration, rolling updates, and health-checks \cite{poniszewska-maranda_kubernetes_2021}. In this way, it's possible to imagine Kubernetes delegating computationally intensive tasks such as data aggregation algorithms or model training to a fleet of powerful containers, which communicate to one or more client containers. Thus, a containerized artifact could be intelligently scaled to accommodate new users or take advantage of computing clusters when they become available. While our main focus remains reproducibility, the modular-first approach of containers suggest substantial downstream benefits in terms of portability and especially scalability. 

Traditionally, one of the largest barriers to embracing the containerization paradigm is the learning curve required to containerize an application. Using a \texttt{Dockerfile}, a developer must specify a container's base image, how to copy application code into the container, what data must be persistently stored, and how the application's environment must be set up in Docker's DSL (Domain Specific Language). For applications which require the orchestration of multiple container's, a \texttt{docker-compose.yml/compose.yaml} file must be specified, a YAML configuration file outlining what containers the application consists of, which data paths are required, and configures communication channels between containers. Thus, containerizing an application requires considerable knowledge of Docker's capabilities, it's domain-specific language, and the specific needs of the application. However, the rise of agentic AI has dramatically simplified devOps workflows through automation, suggesting that AI agents may be capable of containerizing applications semi-independently. Since AI agents are capable of digesting large code-bases and possess an incredibly broad knowledge base of application architectures, this technology is uniquely suited to containerize systems automatically.


\section{Case Study: IML Systems}


\begin{table*}[htbp]
\centering
\caption{An overview of the ten Interactive Machine Learning (IML) systems developed as part of a graduate seminar. These systems facilitate a number of use cases on a myriad of model architectures, including model steering and explainability (XAI) for LLMs, classifiers, and vision models. Further, these systems serve as case studies demonstrating the feasibility, ease of use, and utility of containerizing visualization research artifacts. Despite utilizing different technology stacks, bespoke visual analytics systems can be containerized with minimal performance overhead.  Screenshots of all 10 applications can be seen in Figure~\ref{fig:systems}.}
\label{tab:systems}
\renewcommand{\arraystretch}{0.85} 
\setlength{\tabcolsep}{3pt}        
\footnotesize                       

\begin{tabularx}{\textwidth}{@{} l X X @{}}
\toprule
\textbf{System} & \textbf{Description} & \textbf{Tech Stack (Frontend / Backend / ML)} \\
\midrule
S1: nfl-hof-vis & Bridges the XAI accessibility gap for non-experts via interactive sports dashboard with "what-if" sliders. & HTML/CSS/JS, Vega-lite / Python (FastAPI + Uvicorn) / scikit-learn, PyTorch, SHAP, pyDVL \\
\addlinespace[2pt]
S2: MolExplain & Interactive dashboard for biomedical researchers to build molecules and predict spatial property impacts. & React + Vite / Python (FastAPI + Uvicorn) / XGBoost, SHAP \\
\addlinespace[2pt]
S3: LexiFlow & Combats NLP tool fragmentation by unifying annotation, collaboration, and training in one workspace. & Next.js + React / Python (FastAPI + Uvicorn), SQLite / scikit-learn, PyTorch, Hugging Face \\
\addlinespace[2pt]
S4: WatchNext & Delivers personalized movie recommendations using collaborative and genre filtering with rich media assets. & Streamlit / Python / scikit-learn \\
\addlinespace[2pt]
S5: LLMTuner & Democratizes RLHF via interactive, code-free LLM fine-tuning using slider feedback and quantization. & Streamlit / Python (FastAPI + Uvicorn) / scikit-learn, PyTorch, BitsAndBytes, Hugging Face \\
\addlinespace[2pt]
S6: BirdWatcher & Mitigates background correlations in image models using real-time heatmaps and human-drawn masks. & Streamlit / Python, matplotlib / torchvision \\
\addlinespace[2pt]
S7: Legislative Constraint Explorer & Replaces static voting tables with an interactive dashboard decomposing congressional votes via sliders. & Streamlit / Python / scikit-learn \\
\addlinespace[2pt]
S8: FewShotForge & Builds intuition on demonstration choice impact on LLM predictions via side-by-side prompt comparisons. & React + Vite / Python (FastAPI + Uvicorn) / Groq API, HuggingFace \\
\addlinespace[2pt]
S9: semantic-shift-viz & Helps linguists analyze LM encodings by visualizing 2-D projections of contextual word embeddings. & React / Python (Flask), SQLite, LanceDB / scikit-learn, PyTorch, umap-learn, Hugging Face \\
\addlinespace[2pt]
S10: SCannotate & Resolves scRNA-seq interpretation bottlenecks via model steering, driver-gene interpretability, and annotation. & React + Vite, plotly.js / Python (FastAPI + Uvicorn), Scanpy, AnnData, HDBSCAN / LightGBM, SHAP, scikit-learn \\
\bottomrule
\end{tabularx}
\vspace{-5mm}
\end{table*}

To test this potential, we containerized and deployed 10 novel IML (Interactive Machine Learning) Systems. Screenshots of all 10 applications we containerizerd can be seen in Figure~\ref{fig:systems}.  These systems were created as part of a graduate seminar and cover a variety of different use-cases and interaction models, from LLM-fine tuning, to scRNA-seq bioinformatics model steering. A complete list of systems created, as well a short description and the technologies used in each can be seen in Table \ref{tab:systems}. IML Systems present an especially salient test bed for an AI-assisted containerization workflow due to their inherently modular nature. It's common for IML systems to feature a heavy-duty Python based backend for machine learning computation, and lighter Javascript-based browser frontends. This layered architecture, coupled with conflict-prone nature of Python machine learning libraries, presents an excellent pilot study for visualization system containerization.

\subsection{Methodology}

To containerize each system, we prompted an AI agent to inspect each system's codebase and draft a containerization plan. We used Cursor IDE (Interactive Development Environment) Version 3.4.17, an agent-driven IDE which empowers LLMs (Large Language Models) to read and edit codebases. Cursor uses Retrieval-Augmented Generation (RAG), which searches the codebase and exposes files, functions, and symbols to the LLM when they become relevant. Further, LLMs running in Cursor can execute Bash Commands to test builds and manage files. We used Cursor Composer v2.5 Fast, an LLM based on the Kimi K2.5 checkpoint and fine-tuned for coding tasks. Situated at the root of each project, we gave the agent the following prompt:

\begin{quote}
Review this project and evaluate what backend \& frontend technologies it uses. Come up with a plan to containerize this project using Docker and report back what technologies it uses, and how those technologies inform your containerization plan.
\end{quote}

After perusing each system's codebase, the agent generated a Markdown file which reported on technologies used by the application and it's containerization plan. These markdown files typically included a breakdown of specific languages, librariers, and packages used in the backend and frontend, outlines for the \texttt{Dockerfile}, \texttt{.dockerignore}, \texttt{docker-compose.yml}/\texttt{compose.yaml} files, updates to documentation, obstacles to containerization and potential solutions, and verification checklist. This plan was fact-checked, with a human reviewing that the plan was logical and the technologies enumerated were correct. Occasionally, the agent would ask follow-up questions, related to which services the container should expose, whether to offer a CPU-fallback for GPU-first systems, or whether to optimize the build for development or production. In all cases, the agent was instructed to focus on a production build, offer a CPU-fallback when applicable, and expose the minimum services needed for deployment. Once these follow-ups had been answered and the plan reviewed, the agent was allowed to proceed with its containerization plan.

\subsection{Results \& Experience}



The substantive actions taken by the agent consisted of drafting \texttt{Dockerfile}, \texttt{.dockerignore}, and \texttt{docker-compose.yml}/\texttt{compose.yaml} files, specifying how the application's environment can be created and what services should be exposed. Most applications utilized one-container builds, however, projects with build stages leveraged multiple containers. The automated containerization pipeline demonstrated a high initial success rate, with nine out of ten systems building and executing successfully on the first attempt. However, we note a number of common implications regarding the nature of these systems.

\noindent \textbf{Accessing Specialized System Hardware: } A recurring challenge emerged around hardware acceleration and native GPU access. Three systems: LexiFlow (\textbf{S3}), LLMTuner (\textbf{S5}), and semantic-shift-viz (\textbf{S9}) have backends which heavily utilize on GPU computation for tasks like deep learning inference, dimensional reduction, or intense hardware-accelerated rendering. Initial local containerization efforts failed since a vanilla Docker daemon uses standard container runtimes with no awareness of hardware accelerators like NVIDIA GPUs. Resolving these bottlenecks required configuring the \href{https://hub.docker.com/r/nvidia/cuda}{NVIDIA CUDA Container Toolkit}, a daemon which acts an intermediary between the container manager and execution engine. Furthermore, to maximize the portability of these systems across architectures lacking dedicated hardware, the AI agent engineered automated CPU fallback layers, allowing them to execute gracefully in resource-constrained environments. Since the NVIDIA CUDA Container Toolkit strictly separates the NVIDIA Kernel Driver from user-level libraries such as the CUDA toolkit, portability is maximized since systems may run on both consumer and enterprise cloud server GPUs without recompiling. However, a tradeoff comes in terms of shifting the configuration burden from the application software layer to the underlying systems infrastructure layer.

\noindent \textbf{Isolating Toolchains via Multi-Stage Builds: } A secondary engineering challenge involved managing the architectural tension between frontend and backend tech stacks, exemplified by {MolExplain (\textbf{S2})}, which integrates a reactive Vite web interface with a deep-learning-driven cheminformatics backend. In traditional local development environments, reproducing such a system is notoriously brittle, forcing a researcher to manage node package managers (\texttt{npm}) alongside low-level system-level C/C++ libraries. To enforce strict modularity while ensuring artifact portability, we leveraged Docker's multi-stage builds orchestrated via Docker Compose. In the initial compilation phase, a dedicated Node.js build stage compiles the Vite frontend application. The resulting production assets are then copied over to a backend image, discarding the node compilation footprint. This decoupled runtime environment allowed the isolated declaration of native Linux visualization libraries: \texttt{libxrender1} and \texttt{libxext6} for rendering of molecular graphics, and \texttt{libgomp1} for parallel matrix calculations. Packaging these disjoint requirements into a multi-stage build mitigates the risk of external software decay. However, a trade-off of this design pattern is the heightened complexity of the configuration script, where alterations to build layers can prolong compilation cycles.

\noindent \textbf{Architectural Homogeneity: } Due to the nature of the seminar and the modern pedagogical environment, projects naturally converged onto variants of a Thin Client-Server configuration (\textbf{A1}), with most pairing React or Vite frontends with Python-based backends, or using Streamlit\footnote{\url{https://streamlit.io/}}, a Python-based framework for building web apps. This architectural homogeneity reflects the integration of LLMs and generative AI assistants into the software development pipeline. Generative AI can dramatically accelerate the development of IML systems by providing rapid code scaffolding and acting as a stabilizing force that funnels developers toward well-documented design patterns. Additionally, by leveraging an automated AI agent to analyze these codebases, we were able to synthesize Docker configurations in a mere 30 to 60 minutes per system, mitigating the historic ``learning curve" argument against containerization in academic research. However, this approach comes with the risk of introducing an architectural path dependency that could implicitly disincentivize unconventional system designs.

\section{Discussion}

In visualization research, evaluation and replication has been difficult since specialized research software is inherently tied to specific temporal software ecosystems and relies on platform-specific libraries. When a paper is published, its system becomes a static artifact trapped in screenshots and videos, with significant effort required to run once more on modern hardware. While tools like the Data Visualization Catalogue\footnote{\url{https://datavizcatalogue.com/}} and Data Viz Project\footnote{\url{https://datavizproject.com/}} categorize static chart types, the visualization community lacks an equivalent taxonomy for complex interactive systems. A containerized registry would allow researchers to explore full visual analytics pipelines, analyzing how previous work handled complex state coordination, coordinated multiple views, or pioneered new visualization techniques and algorithms. Interactive systems suffer since it's challenging to run a user study comparing System $A$ directly against System $B$ due to deployment friction. If both systems are containerized and available on a standardized registry, a researcher could deploy them side-by-side on a standardized server, facilitating controlled comparative user studies and baseline benchmarking.

Docker is still a changing platform and there are concerns to be had in regards to the long-term feasibility of Docker containers. It's unclear whether today's containers will be able to run in the far future as hardware and chip architectures continue to evolve. In addition, containerization techniques alone are not sufficient to fully address variable explosion (\textbf{C3}), which can arise from the particular needs of experts in design studies and interpersonal relationships between collaborators \cite{akbaba2023collaboration}. However, adopting a standard software engineering practice like Docker (and the \textit{Dockerfiles} which provide a recipe for an appropriate execution environment) would add a formatted provenance of how the systems were originally run and would improve the chances that they would be able to be run in the future. Containers and their \textit{Dockerfiles}, along with more stable representations like images and videos, could be kept together in repositories to provide the clearest possible picture to researchers of the implementation and intended workflow of VA systems.

In the current landscape, every new system must ``reinvent the wheel", with researchers building custom components for every new project. While containerization presents an approach which may mitigate this duplication, Docker containers are still heavy compared to native applications and less portable than a true virtual-machine solution. Thus, visualization research should shift towards a truly modular paradigm which leverages composable microservices, allowing systems to be assembled like Lego bricks rather than authored as a single monolithic stack. However, technical solutions alone cannot bridge the reproducibility gap; driving long-term change requires institutional incentives and deliberate community mobilization. To this end, we propose a two-pronged initiative aimed at retrofitting historical visual analytics artifacts and standardizing future submissions. First, we plan to deploy a web form to collect, containerize, and preserve legacy visual analytics platforms that are currently at risk of software decay. This ongoing effort will aim to rescue foundational research from obsolescence and provide a dataset for analyzing architectural commonalities across the field. Second, to sustain this momentum into the future, we advocate for the introduction of a specialized \textit{Vis Containerization Stamp}. A domain-specific visual analytics stamp would officially certify that a system's runtime environment is verified, portable, and equipped with stable fallback mechanisms. By formalizing this stamp within major visualization venues, the community can effectively institutionalize artifact preservation.


\section{Conclusion}

In this paper, we have addressed the challenge of the reproducibility crisis in visualization systems research by advocating for the adoption of containerization. We presented a comparison weighing the pros and cons of traditional visualization system, analyzing how each paradigm impacts research utilities and establishing containerization as a highly viable compromise for modern research software. Next, we provided an outline for containerizing visual analytics artifacts, validating this workflow through a case study involving the containerization and public hosting of ten novel visual analytics systems. By continuing to expand this containerized infrastructure, we aim to foster an open, benchmarkable ecosystem that allows future evaluators to conduct rigorous studies and push the boundaries of visualization systems research.


\bibliographystyle{abbrv-doi}

\bibliography{template}

\end{document}